\documentclass[12pt]{iopart}
\usepackage{iopams}  
\usepackage{amssymb}
\usepackage{setstack}  
\usepackage{graphicx}
\usepackage{bm}
\usepackage{amsbsy}
\usepackage{amstext}
\usepackage{iopams}
\usepackage[T1]{fontenc}
\usepackage[latin9]{inputenc}

\newcommand{\Op}[1]{{\boldsymbol{\mathrm{\hat{#1}}}}}

\begin{document}

\title{Parameter estimation in atomic spectroscopy using exceptional points}

\author{Morag Am-Shallem and Ronnie Kosloff}
\address{The Fritz Haber Research Center and The Institute of Chemistry, The Hebrew University, Jerusalem 91904, Israel}

\author{Nimrod Moiseyev}
\address{Schulich Faculty of Chemistry and Faculty of Physics,\\Technion, Haifa 3200008, Israel}


\begin{abstract}
We suggest a method for accurate parameter estimation of atomic systems, 
employing the special properties of the exceptional points.
The non-hermitian degeneracies at the exceptional points emerge 
from the description of the spontaneous emission of atomic system in the framework of an open quantum system,
resulting in a non hermitian quantum master equation.
The method is demonstrated for the atomic spectrum of $S \rightarrow P$ transitions of $~^{85}$Rb and $~^{40}$Ca$~^+$.
\end{abstract}


\section{Introduction: Electronic transitions and spontaneous emission in atomic systems}
\label{sec:intro}

Atomic spectroscopy is unique in its experimental accuracy, able to achieve a dynamical range of precision
of up to 18 significant digits. High-performance frequency standards is 
the technological result of this precision leading to applications  such as network synchronization
and GPS \cite{katori2011optical}. 
This technology is enabled by atomic clocks \cite{ye2015clocks,Matsubara2012direct_coomparison_Ca_Sr}. 
High accuracy has implications in other fields of physics such as
radioastronomy (very-long-baseline
interferometry) \cite{nand2011ultra}, tests of general relativity \cite{chou2010optical}, and particle
physics \cite{blatt2008new}. 
Atomic spectroscopy has been a primary source of fundamental constants \cite{mohr20112010}. 
For example, a small deviation of the Rydberg constant can indicate the radius of the proton \cite{pohl2010size}.

\subsection{Electronic transitions and spontaneous emission}

Atomic spectroscopy is the study of electronic transition in atoms. The spectral lines correspond to Bohr frequencies of the transitions between 
energy levels of the atom. Within this viewpoint  the spectral theory involves calculating the
eigenvalues of the atomic hermitian Hamiltonian. The observed spectrum is then predicted by perturbation theory assuming weak excitation
and knowledge of the transition dipole matrix elements. 

The simple picture of atomic spectroscopy is hampered by the notion that atoms are imbedded in the radiation field.
The primary influence is spontaneous emission and Lamb shifts \cite{chaichian2001hydrogen}.
In principle one can employ quantum field theory and treat the radiation field and the atom using a Hamiltonian description \cite{itzykson2006quantum}.
\begin{equation}
\Op H = \Op{H}_{atom} +\Op{H}_{radiation} + \Op{H}_{interaction}.
\label{eq:QFT_hemil}
\end{equation}
Our aim is to concentrate on the atomic spectra. 
We therefore employ a reduced description where we derive effective equations of motion for the atomic
system by tracing out the radiation field. This is the approach incorporated in open quantum systems.
In this case the reduced dynamics  is described by a non-hermitian generator ${\cal L}$.
We will show that due to non-hermitian degeneracies there is a profound and unexpected influence on the atomic spectrum.

\subsection{The L-GKS equation for spontaneous emission} 
\label{sec:SE}
The phenomena of spontaneous emission (SE) cannot be described by a unitary description, 
such as the Schr\"odinger equation for the wave function, 
or the counterpart Liouville-von-Neumann master equation for density matrices.
Hamiltonian-based approaches incorporate only coherent dynamics. 
Dissipation and dephasing phenomena  are properly described by  the quantum master equation  
\cite{agarwal1970master,cohen1998optical}. 
The general structure of the quantum master equation was introduced by
Lindblad \cite{lindblad1976generators} 
and Gorini, Kossakowski and Sudarshan \cite{gorini1976completely} (L-GKS). 
Based on a mathematical construction they obtained the general structure of
the generator ${\cal L}$ of a completely positive dynamical semigroup. 
The L-GKS master equation (known also as the Lindblad equation) 
adds dissipative terms to the master equation which handles SE:

\begin{equation}
\fl
\frac{\partial\Op{\rho}}{\partial t} 
= 
{\cal L}\Op{\rho} 
= 
-\frac{i}{\hbar} \left[\Op{H},\Op{\rho}\right] 
+ 
\sum\limits _{(a,b)} 
\Gamma_{a\rightarrow b} \left(
\Op{A}_{(a,b)} \, \Op{\rho} \, \Op{A}_{(a,b)}^{\dagger} 
-
\frac{1}{2}
\left[
\Op{A}_{(a,b)}^{\dagger}\Op{A}_{(a,b)}, \Op{\rho} 
\right] _+
\right),
\label{eq:LGKS_schrod}
\end{equation}
where the  $ \left[ \cdot, \cdot \right] $ denotes a commutator,
and the $ \left[ \cdot, \cdot \right] _{+} $ denotes an anti commutator.

The first term is the commutator of the Hamiltonian with the density matrix, 
which generates the unitary dynamics. 
The second term is the dissipator, 
which generates the spontaneous emission.
The sum is over the pairs of levels $(a,b)$: 
Each of the annihilation operators 
\begin{equation}
\Op{A}_{(a,b)} \equiv \Op{A}_{a\rightarrow b}
= 
\left| b \left\rangle \right\langle a \right|
\label{eq:SE_operator}
\end{equation}
generate a decay from the upper source level $\left| a \right\rangle$ 
to the lower destination level $\left| b \right\rangle$. 
The anti commutator 
$ \left[
\Op{A}_{(a,b)}^{\dagger} \Op{A}_{(a,b)}, \Op{\rho} 
\right] _+ $
expresses the decrease in population of the excited state $\left| a \right\rangle$, 
while the resulting increase of population of the lower state $\left| b \right\rangle$
is expressed by the term 
$ \Op{A}_{(a,b)} \, \Op{\rho} \, \Op{A}_{(a,b)}^{\dagger} $.  
Note that the anti commutator contains the term 
\begin{equation}
\Op{A}_{(a,b)}^{\dagger} \Op{A}_{(a,b)} 
=
\left(
\left| a \left\rangle \right\langle b \right|
\right)
\left(
\left| b \left\rangle \right\langle a \right|
\right)
=
\left| a \left\rangle \right\langle a \right|
\equiv
\Op{P}_a,
\end{equation} 
where $\Op{P}_a$ is the projection operator, 
projecting on the subspace spanned by $ \left| a \right\rangle $.
Therefore, the decrease in population of the excited state is expressed using only the population on this state, 
and does not require knowledge of other states.

The decay rate for the pair of levels $(b,a)$, $\Gamma_{a\rightarrow b} $, 
can be obtained by a microscopic derivation of the quantum optical master equation from the Hamiltonian of Eq. (\ref{eq:QFT_hemil}) 
under the weak coupling limit.  
The Born-Markov approximation is employed where the perturbation parameter is the dipole interaction between the atom and the radiation 
at temperature $T=0$.
The rate obtained is equivalent to the golden rule formula 
\cite{breuer2002theory}: 
\begin{equation}
\Gamma_{a\rightarrow b} 
= 
\frac{4}{3}
\frac{\omega_{a  b}^3 }{\hbar c^3} 
\left| d_{a  b} \right| ^2
\end{equation}
with $\omega_{ab}$ as the transition frequency,
$c$  the speed of light, 
and $ d_{a b} $ the transition dipole matrix element.
For states with defined angular momentum, the transition dipole matrix element becomes:
\begin{equation}
\Gamma_{a\to b} = 
\frac{4}{3}
\frac{\omega_{ab}^3}{c^2} \alpha
\frac{\left| \left\langle J_a \left|\left| \Op{r} \right|\right| J_b \right\rangle \right | ^2}{2J_b+1}.
\end{equation}
Here, $\alpha$ is the fine structure constant,
and $J_a$, $J_b$ are the angular momenta of the states 
$\left|a\right\rangle$ and $\left|b\right\rangle$.
$ \left\langle J_a \left|\left| \Op{r} \right|\right| J_b \right\rangle $ is the reduced dipole matrix element between $J_a$ and $J_b$.

The total decay rate from a state $\left| a \right\rangle$ 
is the sum $\Gamma_a = \sum_{ b} \Gamma_{a\rightarrow b}$. 
This decay rate defines the lifetime of the excited state: $\tau_a = \Gamma_a^{-1} $. 

The spontaneous emission rate is completely determined by the fundamental physical constants: 
i.e. magnetic moment of the electron and the nuclei, etc.
These constants determine the values of the energy levels splitting and lifetime.
By inversion, an accurate measurement of the  energies and lifetime constitutes an appropriate determination 
of universal parameters.

\subsection{Population leakage}
\label{sec:pop_leak}

Typically in atomic systems the excitation and de-excitation transitions are not closed.
Population can leak to other levels of the atomic system.
The population is expressed by the diagonal entries in the density matrix $\Op{\rho}$, 
and the total population is $\Tr\lbrace\Op{\rho}\rbrace$.
The dissipative term in Eq. (\ref{eq:LGKS_schrod}) conserves the total population in the system, 
i.e. $\partial_t \Tr\lbrace\Op{\rho}(t)\rbrace = 0$. 
To incorporate population loss 
we utilize the fact that decrease of population in an excited state 
is described by the anti commutator terms, 
which uses only the population on this state 
and does not require knowledge about other states.
Therefore the dissipator ${\cal L}$ will include additional terms 
composed only from the anti commutators. 
Such terms cause a decrease in the population of the excited state 
which are not compensated by an increase of population of other states.
For each excited state $ \left| a \right\rangle $ the additional term will have the form:
\begin{equation}
{\cal L}_{leak}^{(a)} \Op{\rho} 
= 
- \frac{1}{2} 
\Gamma_{a,leak} 
\left[
\Op{P}_a, \Op{\rho} 
\right] _+ .
\label{eq:leakage_term}
\end{equation}
The total decay rate from the state $\left| a \right\rangle$ 
is now $\Gamma_a = \Gamma_{a,leak} + \sum_{ b} \Gamma_{a\rightarrow b}$.
We define $\chi_{a} = \Gamma_{leak}/\Gamma_{a}$ 
as the \emph{branching fraction} that decays from the excited state 
$\left| a \right\rangle$ to states out of the primary system. 
Introducing such leaking terms into the dissipator reduces the total population, 
and therefore  $\partial_t \Tr\lbrace\Op{\rho}(t)\rbrace < 0$. 

The total spontaneous emission part of the L-GKS equation will have the form:
\begin{equation}
\fl
\begin{array}{rl}
{\cal L}_{SE} \Op{\rho} 
&= 
\sum\limits _{a} 
\left(
\sum\limits _{b} 
\Gamma_{a\rightarrow b} 
\left(
\Op{A}_{(a,b)} \, \Op{\rho} \, \Op{A}_{(a,b)}^{\dagger} 
-
\frac{1}{2}
\left[
\Op{P}_a, \Op{\rho} 
\right] _+
\right)
-
\frac{1}{2}
\Gamma_{a,leak} 
\left[
\Op{P}_a, \Op{\rho} 
\right] _+ 
\right)
\\
&= 
\sum\limits _{a} 
\left(
\sum\limits _{b} 
\Gamma_{a\rightarrow b} 
\Op{A}_{(a,b)} \, \Op{\rho} \, \Op{A}_{(a,b)}^{\dagger} 
-
\frac{1}{2}
\Gamma_{a} 
\left[
\Op{P}_a, \Op{\rho} 
\right] _+ 
\right)
.
\end{array}
\label{eq:SE_including_leakage_general}
\end{equation}
If the excited state  $\left| a \right\rangle$  
decays to a manifold $B$ with $N_B$ states $\left| b \right\rangle \in B$
with equal decay rate, 
then we have $\Gamma_{a\rightarrow b} = (1-\chi_a)\Gamma_a /N_B$. 
The spontaneous emission part will have the form:
\begin{equation}
{\cal L}_{SE} \Op{\rho} 
= 
\sum\limits _{a} 
\Gamma_{a} 
\left(
\frac{(1-\chi_a)}{N_B}
\sum\limits _{b \in B} 
\Op{A}_{(a,b)} \, \Op{\rho} \, \Op{A}_{(a,b)}^{\dagger} 
-
\frac{1}{2}
\left[
\Op{P}_a, \Op{\rho} 
\right] _+ 
\right)
.
\label{eq:SE_including_leakage_equal_rate}
\end{equation}

\subsection{Pure dephasing}

Pure dephasing is the loss of coherence without change in population. 
Random fluctuations of the energy levels will generate pure dephasing. 
A possible mechanism is caused by elastic collisions with other atoms in the chamber. 
An additional mechanism is caused by noise in the monitoring or driving laser.
As a result, the pure dephasing rate can be controlled, 
for example by changing the density of the atomic gas, 
or by generating fluctuations in the external field. 
We denote the pure dephasing rate by $\Gamma_{deph}$. 

Within the L-GKS equation pure dephasing is described by a generator ${\cal L}$  which commutes with the Hamiltonian, 
for example ${\cal L}= \Gamma_{deph} \left[\Op H , \left[\Op H, \cdot \right]\right]$

\subsection{The Heisenberg form}

An alternative description is to describe the dynamics in an  operator base. 
As a result the L-GKS equation is employed in the Heisenberg representation 
\cite{alicki2001quantum,breuer2002theory,morag2015threeApproaches},  
the hermitian conjugate of Eq. (\ref{eq:LGKS_schrod}). 
The equation of motion for an operator $\Op{X}$ becomes: 
\begin{equation}
\fl
\frac{d}{d t} \Op{X} 
=
\frac{\partial\Op{X}}{\partial t} 
+ 
\frac{i}{\hbar} \left[\Op{H},\Op{X}\right] 
+ 
\sum\limits _{(a, b)} 
\Gamma_{a\rightarrow b} \left(
\Op{A}_{(a, b)}^{\dagger} \Op{X} \Op{A}_{(a, b)} 
-
\frac{1}{2}
\left[
\Op{P}_{a}, \Op{X} 
\right] _+
\right).
\label{eq:LGKS_heis}
\end{equation}
For system with population leakage the equation will have additional anti commutator terms 
as in Eqs. (\ref{eq:SE_including_leakage_general}) and (\ref{eq:SE_including_leakage_equal_rate}).

\section{Dynamics at the exceptional points and parameter estimation}

The dynamics generated by $\mathcal{L}$ will be represented by an explicit matrix vector notation. 
The density matrix $\Op{\rho}$, which is an element in Liouville space, is represented as a vector, 
while $\mathcal{L}$, which is a linear superoperator operating  in this space,
is represented by a matrix.
There are a few methods to generate such a representation 
cf. a recent demonstration \cite{morag2015threeApproaches}.
In this study we employed the Heisenberg approach for the two-level systems, 
and the vec-ing approach for larger systems.
The vec-ing approach flattens the density matrix into a vector, 
representing the L-GKS generator by an appropriate matrix.
This results in $N^2 \times N^2$ matrices for the L-GKS generator.
We denote the vector representation of the density matrix $\Op{\rho}$ as $\vec \rho$, 
and the matrix representation of $\cal L$ by $L$. 
In this notation, Eq. (\ref{eq:LGKS_schrod}), is expressed by a matrix-vector equation:
\begin{equation}
\dot{\vec \rho} =  L \vec \rho
\label{eq:Ydot_LY}
\end{equation}
The eigenvalues of the matrix $L$ reflect the non-hermitian dynamics generated by ${\cal L}$.
In general they are complex with the steady state eigenvector having an eigenvalue of zero.

\subsection{L-GKS Dynamics and exceptional points}

The solution for  Eq. (\ref{eq:LGKS_schrod}), given an initial density matrix $\Op{\rho}_0$,
and assuming that the generator $\cal{L}$ is time-independent,
can be formally expressed by:
\begin{equation}
\Op{\rho}(t) = e^ { \mathcal{L} t } \Op{\rho}_0.
\end{equation}
In the matrix-vector representation we have:
\begin{equation}
\vec \rho(t) = e^{ L t} \vec \rho(0).
\label{eq:expMt_Y0}
\end{equation}

The dynamics described by Eq. (\ref{eq:expMt_Y0}) typically is described by  a sum of decaying oscillatory exponentials.
The dynamics of expectation values of operators, as well as other correlation functions, will have the analytical form 
(see \ref{sec:appendix:polynomial}): 
\begin{equation}
\left\langle X(t) \right\rangle =
\sum_{k}\;d_k \exp[-i \omega_kt]\;,
\label{eq:har}
\end{equation}
where $-i \omega_k$  are the eigenvalues of $L$, 
$d_k$ are the associated amplitudes, 
and both $\omega_k$ and $d_k$ can be complex.

The spectrum of the non-hermitian matrix $L$ is a function of the external parameters of the system.
For specific values the spectrum becomes  incomplete.
This is due to the coalescence of several eigenvectors, denoted as a non-hermitian degeneracy. 
For such parameters the matrix $L$ is not diagonalizable.
Such points in the parameter space are known as \emph{exceptional points} (\emph{EP}).
At the exceptional point the dynamics  has a polynomial character. 
The temporal value of expectation values of operators has the form: 
\begin{equation}
\left\langle X(t) \right\rangle =
\sum_{k}\sum_{\alpha=0}^{r_{k}}\; {d}_{k,\alpha} t^{\alpha} \exp[-i\omega_{k}^{(r_k)}t]\;,
\label{eq:exh}
\end{equation}
which replaces the form of Eq. (\ref{eq:har}) 
(see \ref{sec:appendix:polynomial}). 

When two eigenvalues of the master equations coalesce into one, a second-order non-hermitian degeneracy is obtained.
We refer to it as a second order exceptional point and denote it with {\em EP2}. 
A third-order non-hermitian degeneracy is denoted by {\em EP3}.
There are points in the parameter space in which $n$ pairs of eigenvectors coalesce, 
each pair coalesces into a distinct eigenvector. 
They will be denoted as \emph{EP2$^{\,n}$}.

\subsection{Identification of \emph{EP}s using the dynamics}
\label{sec:harmonic_inversion}
The analytical form of decaying exponentials, Eq. (\ref{eq:har}), 
is used in harmonic inversion methods to find the frequencies and amplitudes of the time series signal 
\cite{neuhauser1995FDM, mandelshtam2001FDM,main2000threeHI}.
Harmonic inversion methods are widely used for analysis of experiments 
in diverse fields such as NMR spectroscopy \cite{hu199876FDM_noise}, 
Fourier transform mass spectrometry  \cite{martini2014mass_spectrometry}, 
and ultrafast pump-probe molecular spectroscopy 
\cite{gershgoren2001impuldive}. 

However, at exceptional points the analytical form is different: 
Fuchs et al. showed that applying standard harmonic inversion methods, 
which were designed for Eq. (\ref{eq:har}),
to a signal generated by Eq. (\ref{eq:exh}),
leads to divergence of the amplitudes $d_k$ \cite{fuchs2014harmonic}. 
We used the Pad\'e approximant harmonic inversion algorithm presented in
Refs. \cite{main2000threeHI,fuchs2014harmonic}. 
The divergence of the amplitudes $d_k$ in the vicinity of exceptional points is employed to accurately locate them in the parameter space  \cite{fuchs2014harmonic,morag2014EPbloch}. 

\subsection{Parameter estimation using \emph{EP}s}
\label{sec:parameter_estimation}
The ability to accurately locate the \emph{EP}s in the parameter space is used for parameter estimation. 
The procedure is as follows:
\begin{enumerate}
	\item 
	Accurately locate in the parameter space the desired exceptional point by iterating the following steps:
	\begin{enumerate}
		\item Perform an experiment to obtain a time series of a physical observable. 
		\item Obtain the characteristic frequencies and amplitudes of the signal. 
		\item In the parameter space, estimate the direction and distance to the \emph{EP} and determine new parameters for the next iteration.
	\end{enumerate}
	\item
    At the \emph{EP}, 
	invert the relations between the characteristic frequencies and the system parameters to obtain the system parameters. 
\end{enumerate}

The accurate location of the exceptional points, followed by inverting the relations, 
will lead to an accurate parameter estimation. 
This procedure was used to estimate the parameters of the Bloch system from iterations of time series
\cite{morag2014EPbloch}.
This parameter estimation is robust to uncertainties in the location of the \emph{EP}s. 
The noise sensitivity is affected by the harmonic inversion. 
See a short discussion regarding noise in harmonic inversion methods in  \ref{sec:noise_sensitivity}.

\section{Parameter estimation of Effective two-level systems}

\subsection{Closed two-level systems}

Under the influence of polarized driving fields, 
some atomic transitions behave as a closed two-level system. 
An example is the transition between the hyperfine states 
$\left|5^2\textbf{S}_{1/2}, F=3, m_F=3\right\rangle$ 
and
$\left|5^2\textbf{P}_{3/2}, F=4, m_F=4\right\rangle$ 
of the $\mathrm{^{85}Rb}$ atom,
with $\sigma^+$ polarization.
The selection rules impose that all the transitions - stimulated and spontaneous - 
occur only between these states.
The system parameters are:
System frequency of $\omega_s$=384.229241689 THz (assuming no Zeeman splitting), 
decay rate of $\Gamma$=38.117$\times 10^6 s^{-1}$, 
and dipole moment of $\mu$=2.98931 $ea_0$ \cite{Steck2010Rb85}.
We define the detuning between the system frequency $\omega_s$ 
and the electromagnetic field carrier frequency $\omega_s$ as $\Delta=\omega_s-\omega_L$. 
The resonance Rabi frequency is $\Omega_R = -\mu E_0/\hbar$, where $E_0$ is the amplitude of the electromagnetic field. 

We employ the Heisenberg representation to describe the dynamics. 
We define $\left|s\right\rangle$ as the lower state, 
and $\left|p\right\rangle$ as the upper state.
The dynamics are described by the set of operators:
\begin{equation}
\label{eq:TLSoperators}
\begin{array}{rcl}
\Op{X} &\equiv& 
\left|s\right\rangle \left\langle p \right|
+ \left|p\right\rangle \left\langle s \right|
\\
\Op{Y} &\equiv& 
i \left(
\left|s\right\rangle \left\langle p \right|
- \left|p\right\rangle \left\langle s \right|
\right)
\\
\Op{Z} &\equiv& 
\left|p\right\rangle \left\langle p \right|
- \left|s\right\rangle \left\langle s \right|
\\
\Op{I} &\equiv& 
\left|p\right\rangle \left\langle p \right|
+ \left|s\right\rangle \left\langle s \right|
\\
\end{array}.
\end{equation}

We form a four-vector from these operators. 
We write the Heisenberg equations, Eq. (\ref{eq:LGKS_heis}), for the operators in this vector, 
and get a differential equation with a $4 \times 4$ matrix \cite{morag2015threeApproaches}. 
The conservation of population is expressed by $\frac{d}{dt}\Op{I}=0$. 
Therefore we can omit the equation for the operator $\Op{I}$, 
and add an inhomogeneous term instead. 
This results in the Bloch equations \cite{morag2014EPbloch}. 
To find the exceptional points we need only the homogeneous part of the equation,
which is incorporated in the matrix:
\begin{equation}
\label{eq:M_bloch_matrix}
\text{M}=\left(
\begin{array}{ccc}
-\frac{\Gamma}{2} & \Delta & 0\\
-\Delta & -\frac{ \Gamma}{2} & \Omega_R\\
0 & -\Omega_R & - \Gamma
\end{array}
\right).
\end{equation}

The \emph{EP}s of this matrix compose a deltoid-like curve. 
The curve is demonstrated in Figure \ref{fig:leaking_bloch} 
(the $\chi=0$ curve. The other curves in this Figure refer to systems with population leakage 
and will be described below). 
The cusps of this \emph{EP}-curve are identified as \emph{EP3}. 
The accurate location of the \emph{EP}3 can be used to estimate the parameters of 
such systems, 
as described on Section \ref{sec:parameter_estimation} above and in a previous study \cite{morag2014EPbloch}.

\subsection{Two-level systems with population leakage}

The Bloch equation can be extended to include SE that leaks into states that are external to the Hamiltonian, resulting in population loss,
see Section \ref{sec:pop_leak} above.
Here are two examples for such systems:
\begin{itemize}
	\item \emph{Rubidium atom}.
	Consider the TLS composed by the two hyperfine states 
	$\left|5^2\textbf{S}_{1/2}, F=3, m_F=2\right\rangle$ 
	and
	$\left|5^2\textbf{P}_{3/2}, F=3, m_F=3\right\rangle$ 
	of the $\mathrm{^{85}Rb}$ atom \cite{Steck2010Rb85},
	with $\sigma^+$ polarization.
	The selection rules impose stimulated transitions between these states,
	but the excited state, $\left|5^2\textbf{P}_{3/2}, F=3, m_F=3\right\rangle$, 
	decays spontaneously also to other states in the system.
	Under $\sigma^+$ polarization there are no transitions from these other states back to the TLS. 
	Therefore we can treat this system as a TLS with population loss.

	\item \emph{Calcium ion}.
	Consider the TLS composed by the two states 
	$\left|4^2\textbf{S}_{1/2}, m_J=-1/2\right\rangle$ 
	and
	$\left|4^2\textbf{P}_{1/2}, m_J=1/2\right\rangle$ 
	of the $\mathrm{^{40}Ca^{+}}$ ion,
	with $\sigma^+$ polarization 
	(Cf. Section \ref{sec:Ca_H_line} and Figure \ref{fig:Ca_levels} below).
	Again, there are stimulated transitions between these states, 
	but the excited states decay also to 
	$\left|3^2\textbf{D}_{3/2}\right\rangle$ states ($\approx 6.5\%$ of the decay rate)
	and to the state
	$\left|4^2\textbf{S}_{1/2}, m_J=1/2\right\rangle$ 
	($50\%$ of the remaining decay rate). 
	The population on these states does not revert to the TLS 
	\cite{ramm2013branching_fractions_Ca_II}.
\end{itemize}

The Heisenberg equation in this case is 
(cf. Eq. (\ref{eq:SE_including_leakage_equal_rate}) with $N_B=1$
and Eq. (\ref{eq:LGKS_heis}) ):
\begin{equation}
\frac{d}{d t} \Op{O} 
=
\frac{i}{\hbar} \left[\Op{H},\Op{O}\right] 
+ 
\Gamma \left(
(1-\chi)
\Op{S}_+ \Op{O} \Op{S}_- 
-
\frac{1}{2}
\left[
\Op{S}_+ \Op{S}_-, \Op{O} 
\right] _+
\right).
\label{eq:TLS_leak_heis}
\end{equation}
Here $\Gamma$ is the total decay rate of the excited state.  
It is the sum of the decay rate into the lower level $ \left| s \right\rangle $ 
as in Eq. (\ref{eq:LGKS_heis}) above,
and of the decay rate out of the system as in Eq. (\ref{eq:leakage_term}).  
$\chi$ is the branching fraction that decays to states out of the primary system.
$\Op{S}_+\equiv \left| p \left\rangle \right\langle s \right|$
and
$\Op{S}_-\equiv \left| s \left\rangle \right\langle p \right|$ 
are the raising and lowering operators. 
We write the equations for the four operators of Eq. (\ref{eq:TLSoperators}). 
In this case $ \frac{d}{dt} \Op{I} \ne 0 $ and we cannot omit the equation for this operator.  
The resulting set of equations is:
\begin{eqnarray}
\frac{d}{dt}
\left( \begin{array}{c} 
\Op{X}\\ \Op{Y}\\ \Op{Z}\\ \Op{I}\\ 
\end{array} \right) 
=
\left( \begin{array}{cccc}
-\frac{1}{2}\Gamma & \Delta & 0 & 0\\
-\Delta & -\frac{1}{2}\Gamma & \Omega_R & 0\\
0 & -\Omega_R & -\left(1-\frac{\chi}{2} \right) \Gamma &-\left(1-\frac{\chi}{2} \right) \Gamma \\
0 & 0 & -\frac{\chi}{2} \Gamma & -\frac{\chi}{2} \Gamma \\
\end{array} \right)
\left( \begin{array}{c} 
\Op{X}\\ \Op{Y}\\ \Op{Z}\\ \Op{I}\\ 
\end{array} \right) 
\label{eq:leaking_bloch}.
\end{eqnarray}

\begin{figure}[htbp]
\begin{center}
\includegraphics[scale=0.7]{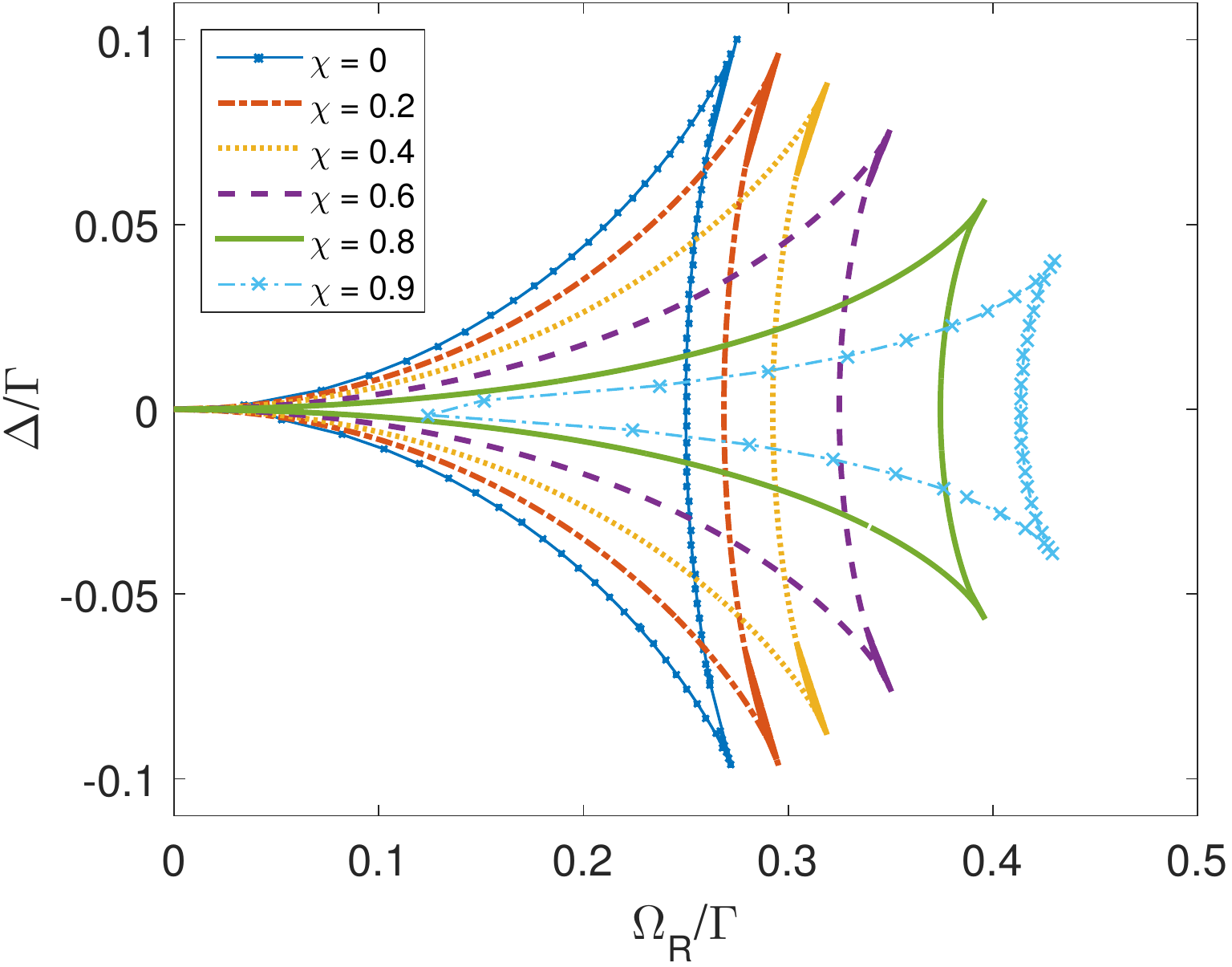} 
\caption{
    A map of the Bloch-like \emph{EP}-curves of the matrix in Eq. (\ref{eq:leaking_bloch}), 
    which describes the dynamics of a two-level system with spontaneous emission,
    when some of the excited population decays out of the system,  
    with $\chi$ as the branching fraction. Figure in scaled coordinates could correspond to 
    any leaking TLS such as Rb or Ca$^+$.
    \\
    The curves are two-fold non-hermitian degeneracy (\emph{EP2}). 
    The curves merge into cusps which are identified as \emph{EP3}. 
    The map of the \emph{EP}-curves can be used for estimation of the system parameters:
    the system frequency, the decay rate, and the branching fraction.
}
\label{fig:leaking_bloch}
\end{center}
\end{figure}
When we look at the exceptional points of this matrix, 
we find that the shape of \emph{EP}-curves is determined by the loss parameter $\chi$. 
Figure \ref{fig:leaking_bloch} shows \emph{EP}-curves for different values of $\chi$. 
The total decay rate $\Gamma$ can be calculated from the eigenvalues of the matrix in Eq. (\ref{eq:leaking_bloch}): the sum of the eigenvalues is always $2 \Gamma$.
The branching fraction $\chi$ of a given system can be found by fitting the resulting \emph{EP}-curve to the appropriate branching fraction.

\section{\emph{EP}s in the H line of the Calcium ion and parameter estimation}

\subsection{The $\mathbf{^{40}}$Ca$^+$ ion}
 
The $^{40}\mathrm{Ca}$ is the most abundant Calcium isotope. 
The total spin of the $^{40}\mathrm{Ca}$ nucleus vanishes.
The ground state of the $^{40}\mathrm{Ca}^{+}$ ion, 
includes 18 electrons in closed shells, 
and the remaining single electron occupying the lower orbital of the 4$^{\textrm{th}}$ shell. 
Therefore $^{40}\mathrm{Ca}^{+}$ ion is isoelectronic to alkali metals. 
However, since the total spin of the nucleus vanishes, there is no hyperfine structure. 

The structure of the energy levels of the $^{40}\mathrm{Ca}^{+}$ ion have been found to be suitable for many applications.	
In particular,  $^{40}\mathrm{Ca}^{+}$ has been used in the field of
quantum computing and quantum information 
\cite{roos2014QIP_ions,barreiro2010experimental_entalglement,    schindler2013QI_trapped_ion,harlander2011trapped_ion_antennae,    Schulz2008microchip_trap_Ca,poschinger2012interaction_laser_qubit_thermal}, 
for atomic clocks and the frequency standard 
\cite{gao2103optical_standard_Ca, gao2015precision_Ca,    Matsubara2012direct_coomparison_Ca_Sr,    ye2015clocks,katori2011optical,gao2015magic_wavelength_Ca,    wolf2011forbidden_clock_transitions}
and recently as a single-atom heat engine
\cite{rossnagel2015heat_engine}.
The spectrum of $^{40}\mathrm{Ca}^{+}$ has been also employed in the quest for drifts of the fine structure constant 
over a time span of many billion years \cite{wolf2008calcium_S_to_P, wolf2008frequency_comb}.

\subsection{The H transition of the $^{40}\mathrm{Ca}^+$ system}
\label{sec:Ca_H_line}

At the ground electronic state the electron occupies the orbital $4\mathrm{s}$, 
with an orbital angular momentum $l=0$.
The total angular momentum including the electron spin becomes $j=\frac{1}{2}$.
The spectroscopic notation for the ion at this state is $4^2\mathrm{S}_{1/2}$.
At the first excited electronic state, the electron occupies the orbital $4\mathrm{p}$, 
with an orbital angular momentum $l=1$.
This state has a fine structure splitting due to spin-orbit coupling either 
 $j=\frac{1}{2}$ (denoted as $4^2\mathrm{P}_{1/2}$), 
 or $j=\frac{3}{2}$ (denoted as $4^2\mathrm{P}_{3/2}$). 

\begin{figure}[htbp]
\begin{center}
\includegraphics[scale=0.5]{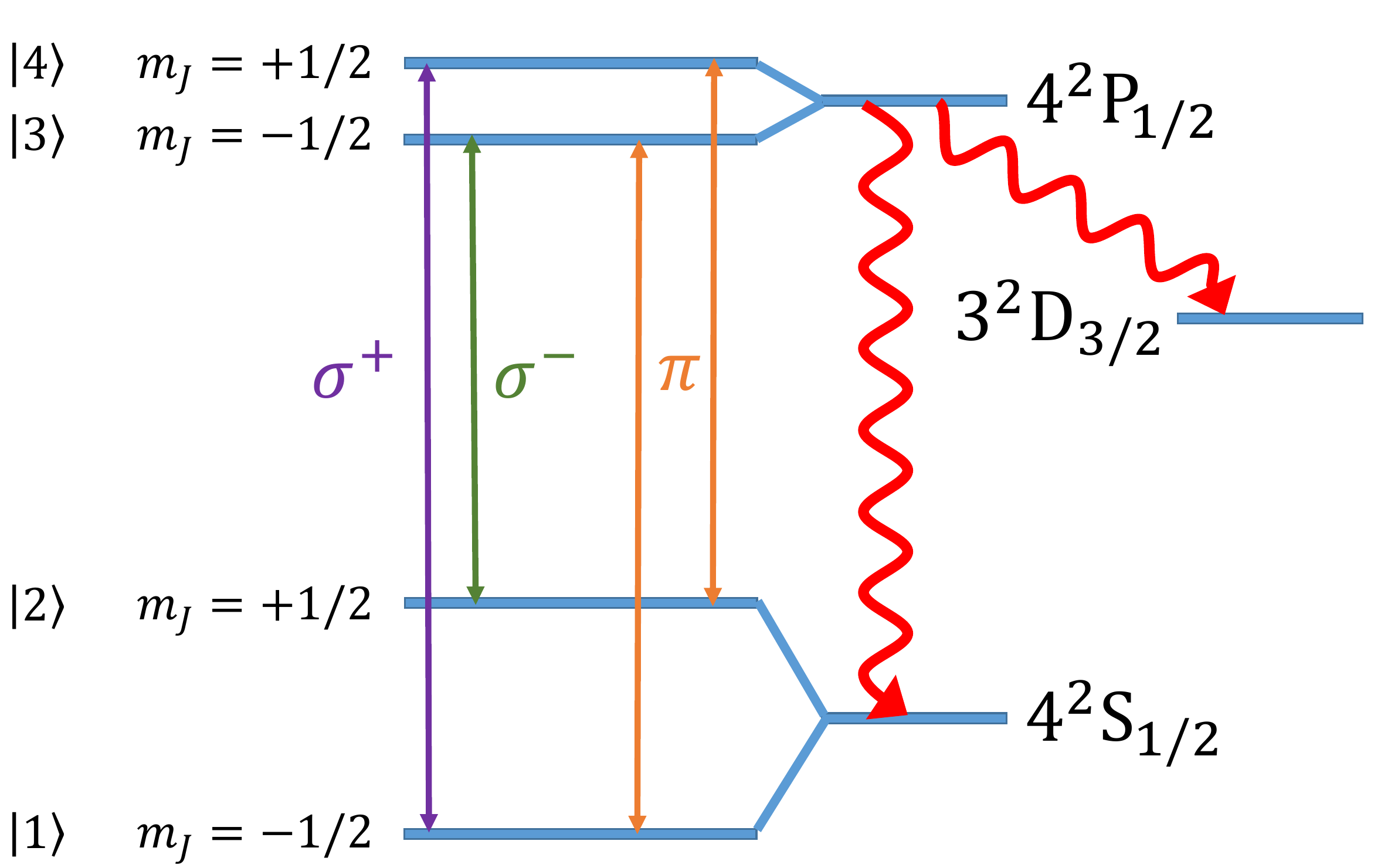}
\caption{
    A scheme of the relevant energy levels in  $\mathrm{^{40}Ca^{+}}$. 
    The $^2$S$_{1/2}$ and $^2$P$_{1/2}$ orbitals have total angular momentum of $j=\frac{1}{2}$. 
    They are split by magnetic field two sub-levels of $m_j \pm \frac{1}{2}$. 
    External electromagnetic fields with $\sigma^+$ and $\sigma^-$ circular polarizations induce 
    $\Delta m = + 1$ and $\Delta m = - 1$ transitions, respectively.  
    Linearly polarized electromagnetic fields ($\pi$ polarization) induce $\Delta m=0$ transitions.
    The excited population at the $^2$P$_{1/2}$ state spontaneously decays
    to the $^2$S$_{1/2}$ and $^2$D$_{3/2}$ states. 
    Energy levels are not to scale.
}
\label{fig:Ca_levels}
\end{center}
\end{figure}

The transition from $4^2\mathrm{S}_{1/2}$ to  $4^2\mathrm{P}_{1/2}$ is known as the H line. 
The transition from $4^2\mathrm{S}_{1/2}$ to  $4^2\mathrm{P}_{3/2}$ is known as the K line. 
These terms stem from the study of the solar spectrum. 
In the following we concentrate on the H line, i.e. the $4^2\mathrm{S}_{1/2}$ $\Leftrightarrow$  $4^2\mathrm{P}_{1/2}$ transition.
The frequency of this transition was measured to be 755222766.2(1.7)MHz \cite{wolf2008calcium_S_to_P}.
The $4^2\mathrm{P}_{1/2}$ has a lifetime of $\tau\approx7ns$, 
and it spontaneously decays back to the $4^2\mathrm{S}_{1/2}$ state, 
as well as to the $3^2\mathrm{D}_{3/2}$ state. 
The branching between these two decays is 
$\Gamma_{\mathrm{P}\to\mathrm{S}} \approx0.935 \times \Gamma_{total}$. 
We treat the decay into the $3^2\mathrm{D}_{3/2}$ state as leakage out of the system, 
with $\chi=1-0.935=0.065$.
Each of the states $4^2\mathrm{S}_{1/2}$ and  $4^2\mathrm{P}_{1/2}$ 
is two-fold degenerated, with sub-levels of $m_j=\pm\frac{1}{2}$. 
When an external magnetic field is applied, 
the Zeeman effect removes degeneracies. 
The magnetic-field-dependent shift in the transition frequency is $\pm$19kHz/$\mu$T  for the $\Delta m=\pm 1$ transitions 
(i.e., the transitions that is induced by circularly polarized electromagnetic fields) \cite{wolf2008calcium_S_to_P}. 
This shift is the sum of two contributions: 
The decrease of energy of the lower sub-level of the $\mathrm{S}_{1/2}$ ($\approx$75\%) 
and the increase of the upper sub-level of the $\mathrm{P}_{1/2}$ ($\approx$25\%). 
The ratio is determined by the appropriate Land\'e factors. 
For a linearly polarized electromagnetic field, $\Delta m=0$ transitions are induced. 
Therefore, we expect to obtain only half of the above shift, i.e. $\pm$9.5kHz/$\mu$T. 
However, in weak magnetic fields this shift is obscured by the natural linewidth
in the standard frequency-domain spectroscopy \cite[for example]{wolf2008calcium_S_to_P}. 
A scheme of the relevant energy levels is presented in Fig. \ref{fig:Ca_levels}. 

\subsection{The system model}

The energy levels structure and the spontaneous emission of the $\mathrm{^{40}Ca^+}$ ion system 
allow the use of  \emph{EP}s in the task of parameter estimation. 
The reduced system  Hamiltonian includes of 4 levels 
(see Figure \ref{fig:Ca_levels} for a sketch of these levels). 
The $4^2\mathrm{S}_{1/2}$ sub levels are denoted as 
$\left|1\right\rangle$ and $\left|2\right\rangle$, 
and the $4^2\mathrm{P}_{1/2}$ sub levels are denoted as 
$\left|3\right\rangle$ and $\left|4\right\rangle$. 
The rotating wave Hamiltonian, 
under the influence of an oscillating electromagnetic field of detuning $\Delta$ and amplitude $\Omega_R$, 
and under a constant magnetic field which induces a split of $\omega_{21}$ between the two $\mathrm{S}_{1/2}$ sub-levels, 
and  a split of $\omega_{43}$ between the two $\mathrm{P}_{1/2}$ sub-levels, 
is: 
\begin{equation}
\label{eq:H_Ca_zeeman}
\Op{H}_0 = \hbar \left(
\begin{array}{cccc} 
\frac{1}{2} (\omega_{43} - \Delta) & 0 & \Omega_R & 0 \\ 
0 & \frac{1}{2} (- \omega_{43} - \Delta)  & 0 & \Omega_R \\ 
\Omega_R & 0 & \frac{1}{2} (\omega_{21} + \Delta)  & 0 \\ 
0 & \Omega_R & 0 & \frac{1}{2} (-\omega_{21} + \Delta) \\ 
\end{array}
\right) 
.
\end{equation}
The spontaneous emission is incorporated into the dynamics by the dissipative part of the L-GKS equation, 
as described in Section \ref{sec:SE} above. 
We used the operators 
\begin{equation}
\Op{A}_{p \to s} 
\equiv 
\left| s \left\rangle \right\langle p \right|,
\end{equation} 
where $\left| s \right\rangle$ denotes the states of the two lower levels - 
$\left|1\right\rangle$ and $\left|2\right\rangle$, 
and $\left| p \right\rangle$ denotes the states of the upper levels - 
$\left|3\right\rangle$ and $\left|4\right\rangle$. 
This results in four terms in the dissipator, 
where each of the $\mathrm{P}_{1/2}$ sub-levels decays to each of the $\mathrm{S}_{1/2}$ sub-levels, 
with rate of 
$\Gamma_{\mathrm{P}_i \to \mathrm{S}_k} 
= 
\frac{1}{2} \Gamma_{(\mathrm{P} \to \mathrm{S})_{total}}
=\frac{1-\chi}{2}  \Gamma_{total}$. 
Another two terms describe the decay from the $\mathrm{P}_{1/2}$ sub-levels to the $\mathrm{D}_{3/2}$ state, 
using only the anti-commutator terms as shown in Eq. (\ref{eq:leakage_term}), 
with the decay rate of $\Gamma_{\mathrm{P}\to\mathrm{D}} = \chi \Gamma_{total}$.
These dynamical terms are incorporated in ${\cal L}$ leading to 
the dynamical equation for the $4\times 4$ density matrix $\Op{\rho}$:
\begin{equation}
\fl
\frac{\partial }{\partial t} {\Op \rho} 
=
{\cal L}\Op{\rho} 
=
-\frac{i}{\hbar}
\left[\Op{H_0},\Op{\rho}\right]
+
\sum\limits _{p \in \mathrm{P}_{1/2}} 
\Gamma_{total} 
\left(
\frac{(1-\chi)}{2}
\sum\limits _{s \in \mathrm{S}_{1/2}} 
\Op{A}_{(p,s)} \, \Op{\rho} \, \Op{A}_{(p,s)}^{\dagger} 
-
\frac{1}{2}
\left[
\Op{P}_p, \Op{\rho} 
\right] _+ 
\right)
.
\label{eq:L_Ca40}
\end{equation}

\subsection{Locations of the \emph{EP}s and parameter estimation}

\begin{figure}[htbp]
\begin{center}
\begin{tabular}{cc}
\includegraphics[scale=0.5]{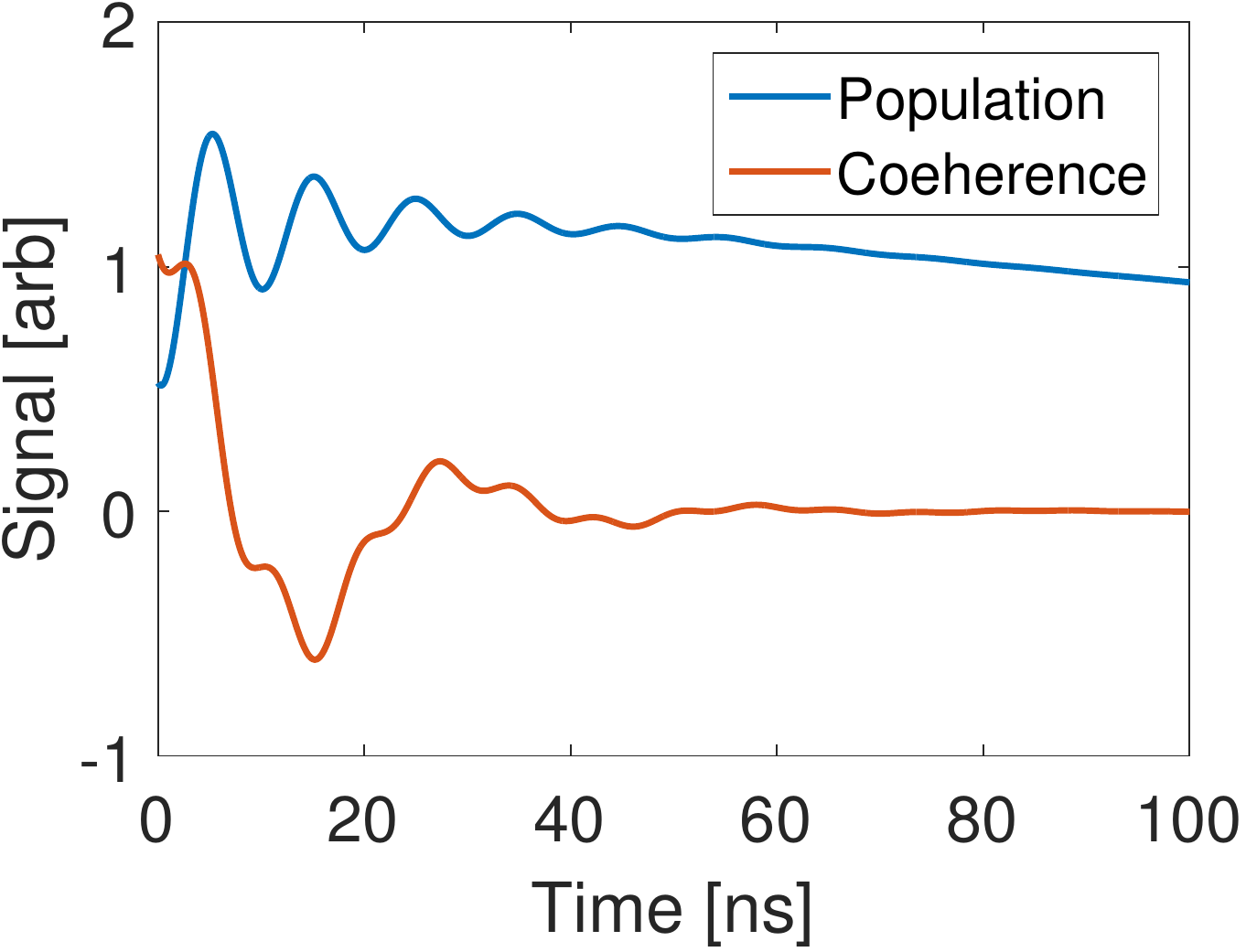} &
\includegraphics[scale=0.51]{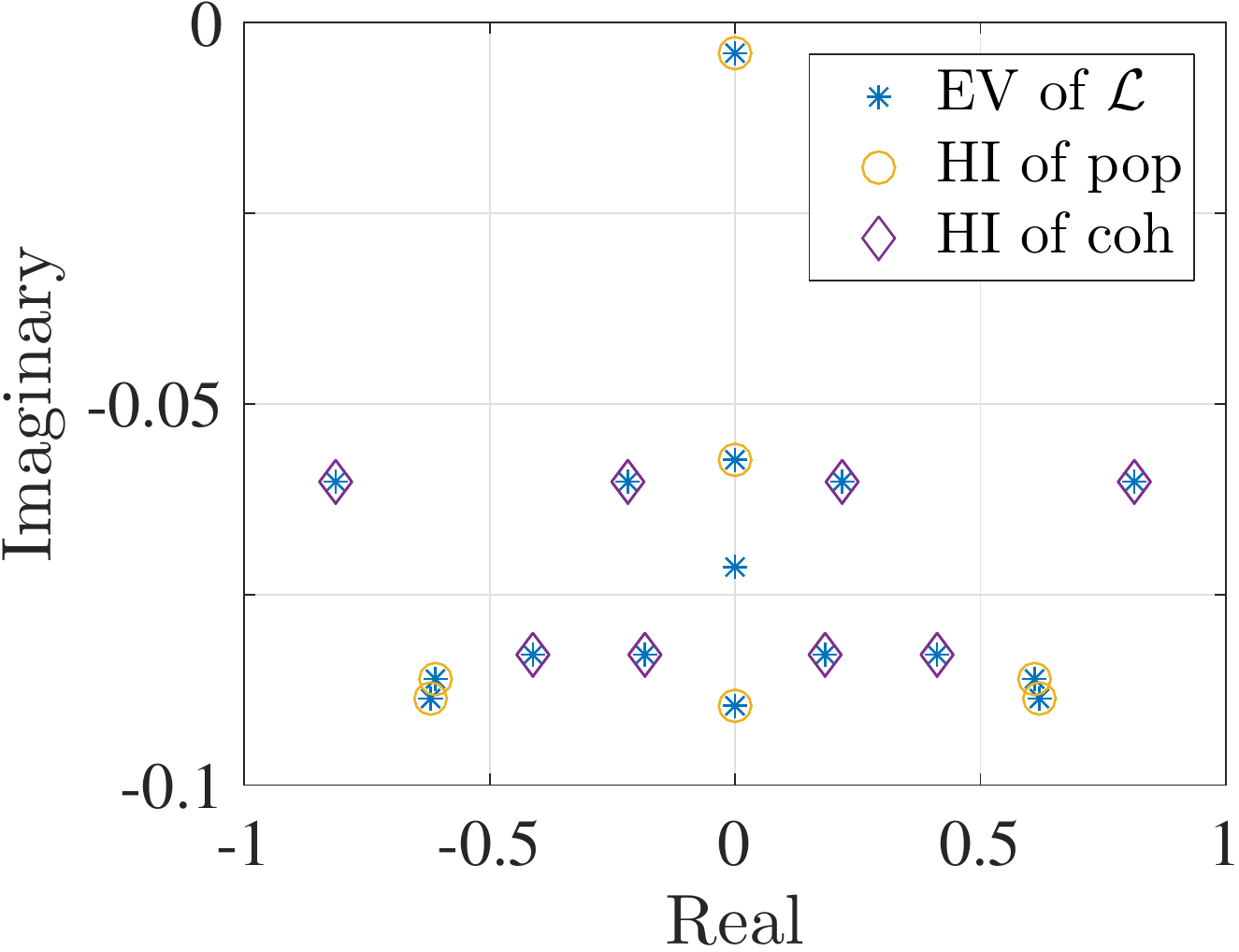} \\
\end{tabular}
\caption{
	Left panel:
	An example of two emission time signals of Ca$^+$, 
   obtained by simulating the dynamics of the populations $\Op{O}_{pop}$ and the coherences $\Op{O}_{coher}$ 
   (see definitions in text). 
	The initial state is the steady state with re-pumping lasers switched on. 
    The transient dynamics is initiated by turning the pumping laser off.
    The time interval in this figure, $100ns$, 
    reflects the population decay life time $\tau_{population} \approx 108ns$.
	Right panel: 
	The locations in the complex plane of the complex frequencies that were obtained from these signals using harmonic inversion (HI). 
	The actual eigenvalues of the generator $\mathcal{L}$ are marked with asterisks. 
	Different subsets of the generator eigenvalues were obtained for different signals.
	The frequencies of the population signal are marked by circles, 
	while the frequencies of the coherence signal are marked by diamonds.
}
\label{fig:time_signal_HI}
\end{center}
\end{figure}

Our task is to find the exceptional points, 
which are the non-hermitian degeneracies of the L-GKS generator of Eq. (\ref{eq:L_Ca40}). 
Experimentally, the first step is to obtain a time series from the driven system.
As an example, we mimic a possible experiment by simulating the time series of the emission signal by solving Eq. (\ref{eq:L_Ca40}). 
The initial condition is obtained by first setting  the laser detuning and amplitude to obtain steady state.
To overcome the population leakage, 
the population from  D$_{3/2}$ is repumped to P$_{1/2}$ using an auxiliary laser.
After a steady state is reached, the auxiliary laser is turned off, obtaining $\Op \rho (0)$.
The decay signal is now collected from $\Op \rho(t)$ for a particular observable in an ordered time grid. 
Two examples for such measurable observables are the populations of the excited states: 
$ 
\Op{O}_{pop} 
\equiv  
\left| 3 \left\rangle \right\langle 3 \right| + 
\left| 4 \left\rangle \right\langle 4 \right|
$,
and the coherences between these states:
$ 
\Op{O}_{coher} 
\equiv  
\left| 3 \left\rangle \right\langle 4 \right| + 
\left| 4 \left\rangle \right\langle 3 \right|
$. 
The left panel of Figure \ref{fig:time_signal_HI} shows an example for such time signals. 
The time series is the input for the harmonic inversion, 
which extracts the frequencies and amplitudes of the time signal. 
The frequencies are determined by the system parameters, 
while the initial state $\Op \rho (0)$ determines the amplitudes. 
The right panel of Figure \ref{fig:time_signal_HI} shows the obtained frequencies in the complex plane. 
The time interval in this figure is $100ns$, 
reflecting the population decay life time $\tau_{population} = \tau_{SE}/\chi \approx 108ns$. 
To map the \emph{EP} at the parameter space, 
this procedure is repeated for other values of the laser detuning and amplitude. 

For any such parameter set defining  $\mathcal{L}$, 
the sum of the 16 eigenvalues of $\mathcal{L}$ can be shown to be: 
$ \sum \limits_{k=1}^{16} \omega_k = 8 \Gamma_{total} $. 
A similar relationship was obtained for the two-level system, 
where the sum of the 4 eigenvalues is $2 \Gamma$. 

\begin{figure}[htbp]
\begin{center}
\includegraphics[scale=0.7]{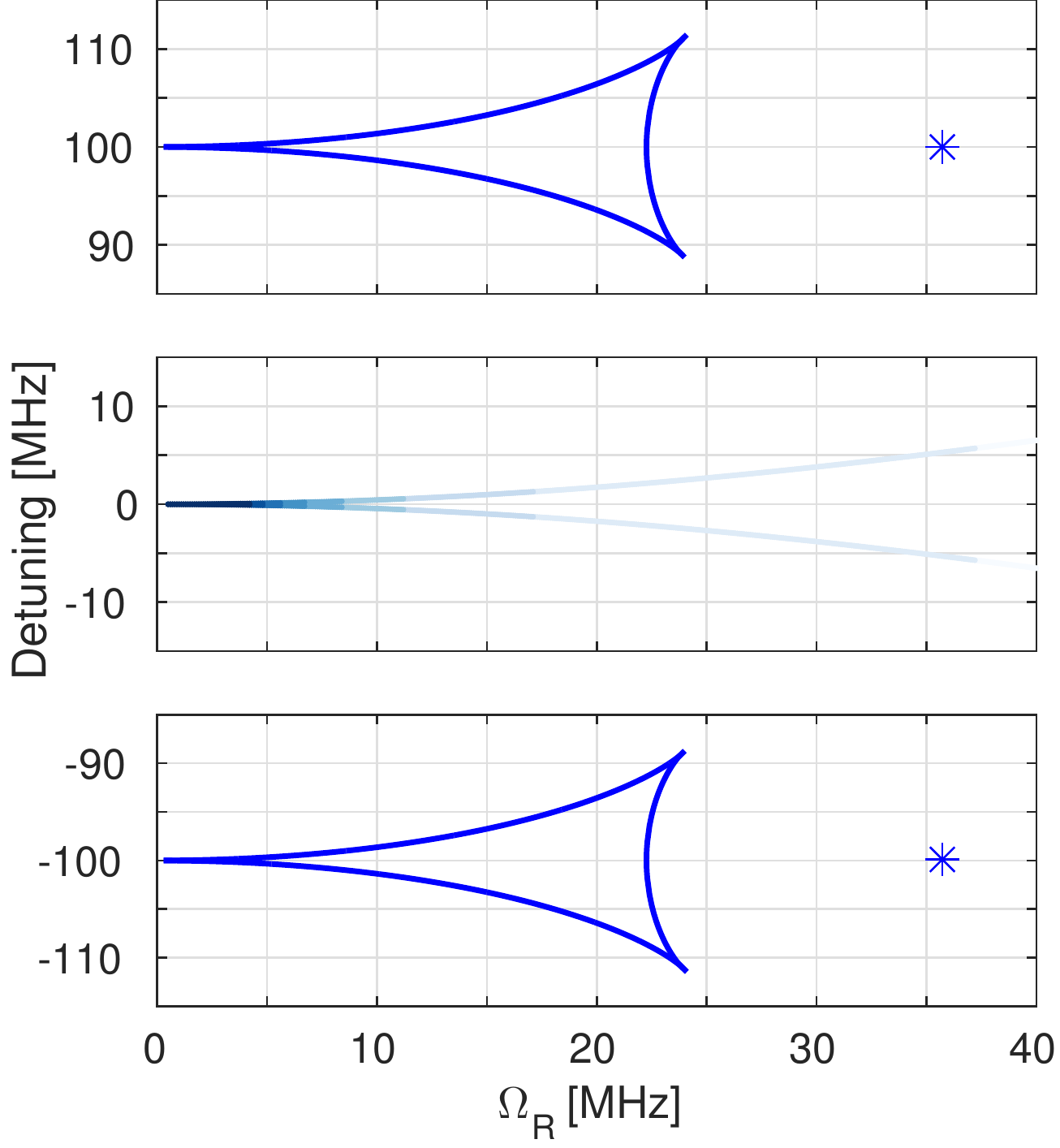} 
\caption{
    A map of the \emph{EP}-curves of $^{40}\mathrm{Ca}^+$ ion (Eq. (\ref{eq:L_Ca40})), 
    with Zeeman splitting of 200MHz, 
    under linearly polarized driving field.  
    Each of the Bloch-like curves (compare to the Bloch curves at Figure \ref{fig:leaking_bloch}) 
    is found on a resonance between a pair of sub-levels, 
    one from $4^2$S$_{1/2}$ and one from $4^2$P$_{1/2}$.
    Note the gaps in the Y-axis of the different curves.
    \\
    To the right of each of the Bloch-like \emph{EP}-curves, there is a point of \emph{EP2$^{\,4}$} (marked with asterisks), 
    in which 4 pairs of eigenvectors coalesce into 4 distinct eigenvectors. 
    The detuning at these points is the splitting of the relevant resonance.
    The amplitude is $\Omega_R=\frac{1}{4}\Gamma_{total}$.
    \\
    Between the two Bloch-like \emph{EP}-curves, 
    at the detuning $\Delta=0$, which is the H-line transition frequency,
    there is a degeneracy-curve of the L-GKS generator.
    It is not decisive whether this curve is an \emph{EP}-curve.
    }
\label{fig:Ca_EP_200}
\end{center}
\end{figure}

The task is to calculate the expected locations of the exceptional points of Eq. (\ref{eq:L_Ca40}) 
in the parameter space of the amplitude and detuning. 
We used the MFRD and the eigenvalues condition number methods for this task
(see \ref{sec:appendix:searchEP} 
and Refs. 
\cite{jarlebring2011doubleEigenvalue,chatelin2011spectralApprox}).
The map of \emph{EP}-curves is shown in Figure \ref{fig:Ca_EP_200}  for the $^{40}\mathrm{Ca}^+$ ion, with Zeeman splitting of 200MHz. 
Note the gaps in the Y-axis.
The resulting \emph{EP}-curves of the 4-level system are more involved than the 2-level case. 
In principle, there could be very high order non-hermitian degeneracies of Eq. (\ref{eq:L_Ca40}). 
In practice, we found few \emph{EP}-curves of second order degeneracy, 
and two other fourth-order \emph{EP}s. 

Close to each of the resonances between the upper and lower levels, 
there is an \emph{EP}-curve which is similar to the deltoid \emph{EP}-curve we got for the Bloch system 
\cite{morag2014EPbloch}.
The exact frequency of the resonances can be found by locating pairs of \emph{EP}s 
with detunings above and below the resonance, 
while maintaining a fixed amplitude. 
The shape of the curves can be fitted to estimate the branching ratio.

These resonance frequencies can be verified by locating the distinct \emph{EP}s
on the right of the Bloch-like curves
(see Figure \ref{fig:Ca_EP_200}).  
These two isolated points are classified as \emph{EP2$^{\,4}$}, 
i.e coalescence of four pairs of eigenvectors with four distinct eigenvalues.
Each of these points 
is located with detuning $\Delta$ at the same frequency as the resonance, 
and amplitude of $\Omega_R=\frac{1}{4} \Gamma_{total}$.
These points can be used also to extract the total decay rate $\Gamma$.

Between the two Bloch-like \emph{EP}-curves, 
in the $4^2$S$_{1/2}$ $\Leftrightarrow$ $4^2$P$_{1/2}$ transition frequency, 
there is a curve of degeneracy points.
However, we could not determine whether these degeneracies are exceptional points.
Anyway, locating these degeneracies can be employed for determining the transition frequency.

To summarize, the suggested procedure for parameter estimation which include four transition frequencies, laser driving power,
spontaneous emission rate and leakage.
\begin{enumerate}
    \item 
    The sum of the 16 frequencies obtained by the harmonic inversion of the time signal can be used to estimate the total spontaneous emission rate:
    $\sum \limits_{k=1}^{16} \omega_k = 8 \Gamma_{total} $.
    \item 
    The locations of the Bloch-like curves are used for the estimation of the frequencies of the resonances between the Zeeman sub-levels. 
    \item 
    The shapes of the Bloch-like curves are used for the estimation of the branching ratio (In particular the EP3 points). 
    \item 
    The locations of the \emph{EP2$^{\,4}$} points are used to verify the resonances frequencies and the decay rate.
    \item
    The location of the degeneracy curve between the Bloch-like curve is used to estimate the H line transition frequency.
\end{enumerate}

Repeating this procedure for various magnitudes of the external magnetic field 
can be used for tracing the Zeeman and Paschen-Back effects.

\begin{figure}[htbp]
\begin{center}
    \includegraphics[scale=0.7]{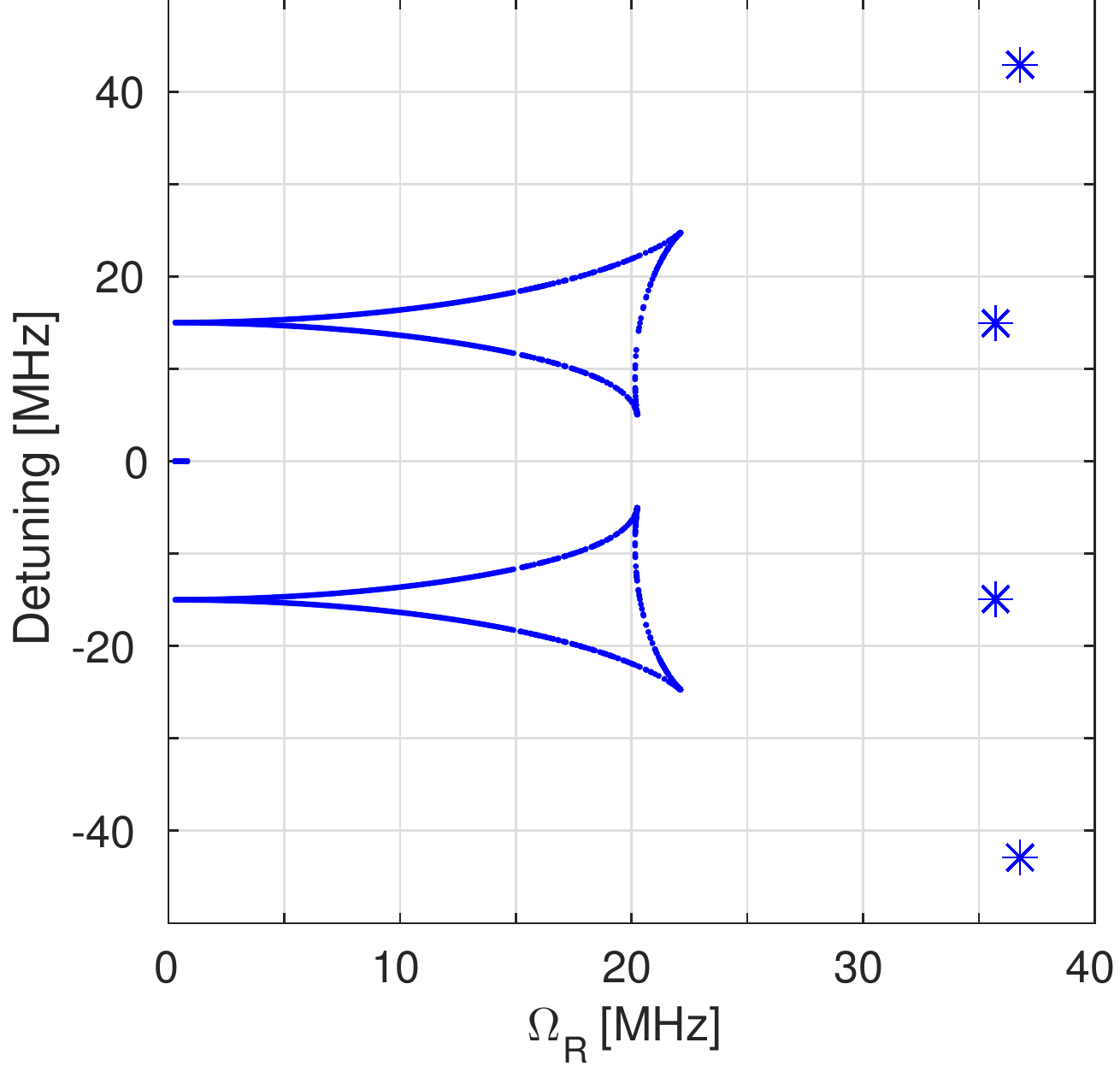} 
\caption{
    A map of the \emph{EP}-curves of $^{40}\mathrm{Ca}^+$ ion (Eq. (\ref{eq:L_Ca40})), 
    with Zeeman splitting of 30MHz. 
    The general structure is similar to the case of 200MHz splitting 
    (Figure \ref{fig:Ca_EP_200})
    but the Bloch-like curves get closer and interfere.
    The interference leads to skewing of these curves.
    The \emph{EP2$^{\,2}$} still can be located and employed for parameter estimation.
    Another two isolated \emph{EP2$^{\,2}$} can be seen at the right corners.
    }
\label{fig:Ca_EP_30}
\end{center}
\end{figure}

For small external magnetic field, the Bloch-like curves approach each other and interfere. 
The shape of these curves is then skewed. 
This is demonstrated in Figure \ref{fig:Ca_EP_30}, 
which shows the map of \emph{EP}-curves for the $^{40}\mathrm{Ca}^+$ ion, 
with Zeeman splitting of 30MHz.
However, the resonances frequencies can be estimated using the locations of the \emph{EP}s at small amplitudes, 
and verified by the location of the isolated \emph{EP2$^{\,4}$} points.
In addition to those \emph{EP}-curves and points, 
we observed two other isolated \emph{EP2$^{\,2}$}, 
at larger detuning and slightly larger amplitude.
We did not find exact analytical expressions for these points.

\subsection{Dependency of the \emph{EP}s on other dephasing rates}

Most of the sources for pure dephasing of laser-driven atomic spectroscopy
are well controlled experimentally, for example varying
the density of the ion gas or the medium gas, 
or the instrument noise in the laser amplitude and frequency. 
Care must be taken 
when analyzing \emph{EP}-curves in atomic spectra to get the relaxation rate $\Gamma$, 
since the rate depends on the various relaxation and dephasing rates in the system.
For example, in the Bloch equations, if the spontaneous emission rate is $\Gamma_{SE}$ and pure dephasing rate $\Gamma_{PD}$,
then the relaxation rate that appears in the matrix of Eq. (\ref{eq:M_bloch_matrix}) is: 
$\Gamma = \Gamma_{SE} - \Gamma_{PD}$ 
\cite{morag2014EPbloch}. 

Generally, every noise source that can be added to the L-GKS equation, 
is reflected by the complex eigenvalues of the generator $\mathcal{L}$. 
These eigenvalues are complex frequencies of the time signal.
Therefore the noise source can be traced by the harmonic inversion. 
The noise will result in changes in the \emph{EP}-curves map. 
The experimental noise will influence only the harmonic inversion. 
See \ref{sec:noise_sensitivity} for a short discussion regarding noise in harmonic inversion methods.

As an example, the influence of noise in the amplitude of the driving laser was analyzed. 
Such a noise is modeled in the L-GKS equation by a double commutator with the laser amplitude operator $\Op{V}_{deph}$, which commutes with $\Omega_R$ amplitude part of the Hamiltonian:
\begin{equation}
\label{eq:V_amplitude_noise}
\Op{V}_{deph} = 
\sqrt{\frac{1}{2}}
\left(
\begin{array}{rrrr}
0 &  0 & 1 & 0 \\ 
0 &  0 & 0 & 1 \\ 
1 &  0 & 0 & 0 \\ 
0 &  1 & 0 & 0 \\ 
\end{array} 
\right).
\end{equation}
The dissipation generated by the double commutator with this operator is not pure dephasing 
since it also generates relaxation.

\begin{figure}[htbp]
\begin{center}
\includegraphics[scale=0.6]{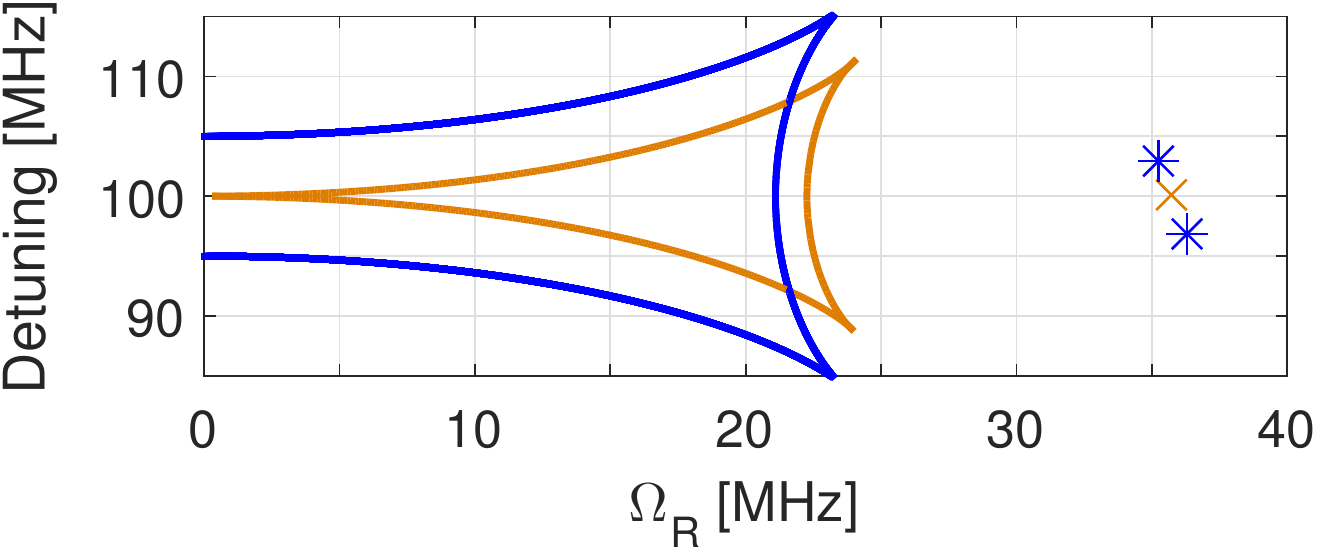}
\caption{
   A map of the upper \emph{EP}-curve of the driven $^{40}\mathrm{Ca}^+$ ion 
   with Zeeman splitting of 200MHz
   and noise in the laser amplitude
   (Eq. (\ref{eq:L_Ca40}) with additional double commutator with $\Op{V}_{deph}$ term, 
   defined in Eq. (\ref{eq:V_amplitude_noise})). 
   Blue line: 
   The \emph{EP}-curve for dissipation rate 
   of $\Gamma_{deph} = 0.01\, ns^{-1} $, 
   along with the associated two \emph{EP2$^{\,2}$} (blue asterisks). 
   The orange dashed line and 'x' are the \emph{EP}-curve 
   and the \emph{EP2$^{\,4}$} obtained
   for the noiseless case ($\Gamma_{deph} = 0\,ns^{-1} $), 
   shown on Figure \ref{fig:Ca_EP_200} above.
   For the case of $\Gamma_{deph} = 0.01\,ns^{-1} $, 
   the two branches do not merge at a small amplitude. 
   Instead, they are split symmetrically around the resonance. 
   The splitting magnitude is equal to the depahsing rate $\Gamma_{deph}$. 
   In addition, the added dephasing splits the \emph{EP2$^{\,4}$} of the noiseless case 
   into two distinct \emph{EP2$^{\,2}$}s.
   }
\label{fig:amplitude_noise_deph}
\end{center}
\end{figure}

We calculated the \emph{EP}-map in the parameter space for dissipation rate of $\Gamma_{deph} = 0.01\, ns^{-1} $ 
with Zeeman splitting of 200MHz. 
The upper Bloch-like \emph{EP}-curve of the results is presented in Figure \ref{fig:amplitude_noise_deph}, 
along with the associated two \emph{EP2$^{\,2}$}. 
For comparison, the upper curve of the noiseless case 
(presented in Figure \ref{fig:Ca_EP_200}) 
is also shown. 
Two prominent differences can be found. 
The first is that the two branches do not merge at a small amplitude. 
Instead they are split symmetrically around the resonance. 
The splitting magnitude is equal to the depahsing rate $\Gamma_{deph}$. 
The second difference is the splitting of the \emph{EP2$^{\,4}$} into two distinct \emph{EP2$^{\,2}$}s.
This splitting is not symmetric, therefore we cannot deduce the system parameters from the locations of these \emph{EP2$^{\,2}$}s.

\section{Discussion}

The irreversible character of the L-GKS equation is well known 
and indicated by the semi-group character of the evolution operator 
\cite{alicki2002invitation,alicki2007qds,breuer2002theory,Nielsen2011quantumComputationInformation}.
The generator of the dynamics ${\cal L}$ is therefore non hermitian. 
This means that non-hermitian degeneracies \emph{EP} play an important role in open quantum systems.

So far, \emph{EP} were studied in the coalescence of two resonances. 
The resonances were metastable states associated with
predissociation or autoionization phenomena and with leaking modes in
waveguides \cite{muller2008EP,moiseyev2011non}.
A theoretical quest for multiple \emph{EP}s \cite{ryu2012multipleEP},
or for high order \emph{EP}s in dissipative physical systems is pursued
\cite{heiss2008chirality,demange2012signature,heiss2015ep3,heiss2015green}, 
specifically in the spectra of atoms in external fields \cite{Cartarius2009epAtomic}. 

The first study  of \emph{EP} in the context of the L-GKS equation, was for the simple two-level-system
described by the Bloch equations  \cite{morag2014EPbloch}. 
In the present study, we generalize to open 
two-level system where population can leak out. 
Then we extend to a four-level system where the splitting can be controlled
by a magnetic field. 
We found a rich and fascinating structure of \emph{EP}'s and \emph{EP}-curves, including higher-order \emph{EP}s.
Such phenomena is expected for many other open quantum systems described by the L-GKS equation.

The methods developed pave the way for a generic framework of employing \emph{EP}s 
for parameter estimation of atomic systems. 
The dynamics near the \emph{EP}s have enhanced sensitivity
due to their analytic properties: 
Small changes in the parameters lead to different harmonic inversion.
Therefore, for parameter estimation, 
the harmonic inversion at the \emph{EP}s is superior to standard inversion methods.

The first stage is to predict the \emph{EP} map of the system: 
The state of an atom driven by a CW laser
can be described in the rotating frame by a time-independent L-GKS equation. 
The parameter space for such a L-GKS generator contains the field amplitude and the detuning frequency.
Such a parameter space can be scanned using the MFRD method to find approximate locations of degeneracies of the generator. 
The location and character of these degeneracies are then examined using the condition number of the eigenvalues, 
to identify and locate the \emph{EP}s. 
The second stage is to search for the predicted \emph{EP}s experimentally: 
The time signals obtained from the experiments are analyzed using harmonic inversion. 
The resulting frequencies and amplitudes are then used to find the degeneracies and exceptional points. 
Finally, we estimate the system parameters by comparing the predicted and the experimental \emph{EP}s.

An interesting different system for an \emph{EP}s search can be two molecular electronic surfaces, with vibrational relaxation.
A simple model for such a system can include only four levels \cite{morag_scaling_2014}, or even three - 
one level from the ground state and two vibrational levels from the excited state.
Such systems can have multiple steady states, and therefore can possess richer dynamics.

\subsection*{Aknowledgements}
We thank Ido Schaefer, Amikam Levy, and Raam Uzdin for fruitful discussions.
We want to thank A. L. Wolf, W. Ubachs, and K. S. E. Eikema for their help on the Ca$^+$ system.
We thank P. Gould for his advice. 
Work supported by the Israel Science Foundation   Grants No. 2244/14 and No. 298/11  
and  by I-Core: the Israeli Excellence Center ``Circle of Light''.

\appendix

\section{Dynamical signature of the \emph{EP}s}
\label{sec:appendix:polynomial}

The solution for the L-GKS equation in the matrix-vector representation, Eq. (\ref{eq:Ydot_LY}), is:
\begin{equation}
\vec \rho(t) = e^{ L t} \vec \rho(0).
\label{eq:aapendix_expMt_Y0}
\end{equation}

When $ L$ is diagonalizable, we can write $ L = T \Lambda T^{-1}$,
for a non-singular matrix $T$ and a diagonal matrix $\Lambda$, 
which has the eigenvalues $\lbrace \lambda_i \rbrace$ on the diagonal. 
Then we have
\begin{equation}
e^{ M t} = T e^{\Lambda t} T ^{-1}.
\end{equation}
The matrix $e^{\Lambda t}$ is a diagonal matrix, which has the exponential of the eigenvalues, $\exp[\lambda_i t]$, on its diagonal.
The resulting dynamics of expectation values of operators, as well as other correlation functions, 
follows a sum of decaying oscillatory exponentials.
The analytical form of such dynamics is:
\begin{equation}
\left\langle X(t) \right\rangle =
\sum_{k}\;d_k \exp[-i \omega_kt]\;,
\label{eq:appendix_har}
\end{equation}
where $-i \omega_k$  are the eigenvalues of $ L$, 
$d_k$ are the associated amplitudes, 
and both $\omega_k$ and $d_k$ can be complex.

For special values of the system parameters
the spectrum of the non-hermitian matrix $ L$ is incomplete.
This is due to the coalescence of several eigenvectors, referred to as a non-hermitian degeneracy.
The difference between hermitian degeneracy and non-hermitian degeneracy is essential: 
In the hermitian degeneracy, several different orthogonal eigenvectors  are associated with the same eigenvalue.
In the case of non-hermitian degeneracy several orthogonal eigenvectors coalesce to a single  eigenvector
\cite{moiseyev2011non}.
As a result, the matrix $ L$ is not diagonalizable, 
and the exponential $e^{ L t}$ cannot be expressed using the eigenvalue decomposition. 

The exponential of a non-diagonalizable matrix $ L$ can be expressed using its Jordan normal form: 
$ L = T J T ^{-1}$. 
Here, $J$ is a Jordan-blocks matrix which has (at least) one non-diagonal Jordan block;
$ J_i = \lambda_i I + N $, 
where $I$ is the identity  and $N$ is has ones on its first upper off-diagonal. 
The exponential of $ L$ is expressed as 
\begin{equation}
e^{ L t} = T e^{J t} T ^{-1}.
\end{equation}
The exponential of the block $J_i$ in $e^{J t}$ will have the form:
\begin{equation}
e^{J_i t} = e^{\lambda_i I t + N t} = e^{\lambda_it} e^{N t}.
\end{equation}
The matrix $N$ is nilpotent and therefore the Taylor series of $e^{Nt}$ is finite, 
resulting in a polynomial in the matrix $Nt$.
This gives rise to a polynomial behaviour of the solution, 
and the dynamics of expectation values of operators will have the analytical form of 
\begin{equation}
\left\langle X(t) \right\rangle =
\sum_{k}\sum_{\alpha=0}^{r_{k}}\; {d}_{k,\alpha} t^{\alpha} \exp[-i\omega_{k}^{(r_k)}t]\;,
\label{eq:aapendix_exh}
\end{equation}
instead of the form of Eq. (\ref{eq:appendix_har}).
Here, $\omega_k^{(r_k)}$ denotes a frequency with multiplicity of $r_k+1$.
Note that for non-degenerate frequencies, i.e. $r_k=0$, 
we have $d_{k,0}=d_k$ and $\omega_k^{(0)}= \omega_k$.
The difference in the analytic behaviour of the dynamics results in non-Lorentzian line shapes,
with higher order poles in the complex spectral domain.
The point in the spectrum where the  eigenvectors coalesce is known as an \emph{exceptional point} ($EP$).

\section{Searching for \emph{EP}s at the parameter space}
\label{sec:appendix:searchEP}
Given a parameters-dependent matrix, 
the task is to find the exceptional points, 
i.e., to calculate the parameters set for which the matrix is not diagonalizable.

\subsection{Condition number of an eigenvalue}
\label{sec:condeig}
The diagonalization of matrices in the vicinity of a defective matrix is extremely sensitive to perturbations. 
The sensitivity of the diagonalization can be characterized by the condition numbers of its eigenvalues.
Therefore the divergence of the condition number of an eigenvalue can be used to find exceptional points.
The condition number of an eigenvalue $\lambda$ of a matrix $A$ with $y$ and $x$ as the corresponding (normalized) left and right eigenvectors, respectively, is defined by:
\begin{equation}
\kappa(\lambda,A) = \frac{1}{y^H x},
\end{equation}
where $y^H$ is the hermitian transpose of $y$ 
\cite{golub1989matrix,moler2004numericalMatlab,chatelin2011spectralApprox}. 
At exceptional points the left and right eigenvectors are perpendicular, 
and the scalar product $y^H x$ vanishes, leading to divergence of the eigenvalue condition number. 
The condition number of the eigenvalues is implemented in the Matlab function CONDEIG. 

\subsection{Newton methods}
There are a few methods that use the special properties of the exceptional points in order to find them iteratively: 
\begin{itemize}
\item
Mailybaev developed a Newton method of finding multiple eigenvalues with one Jordan block and corresponding generalized eigenvectors for matrices dependent on parameters. 
The method computes the nearest value of a parameter vector with a matrix having a multiple eigenvalue of given multiplicity 
\cite{Mailybaev2006computation}.
This method worked well for us in some cases, but failed to find points in which two different eigenvalues had double multiplicity.
\item
Akinola and coworkers used an implicit determinant method to obtain a numerical technique for the calculation of a two-dimensional Jordan block in a parameter-dependent matrix
\cite{Akinola2014jordanBlock}.
\end{itemize}

\subsection{The MFRD method for finding a double eigenvalue of a parameter-dependent matrix}
\label{sec:MFRD}
Jarlebring and coworkers suggested a method that for a given two $n \times n$ matrices, $A$ and $B$, 
computes all pairs ($\lambda$,$\mu$) such that $\lambda$ is a double eigenvalue of $A + \mu B$
\cite{jarlebring2011doubleEigenvalue}.
The method they suggest is the method of fixed relative distance (MFRD).
It is based on the assumption that in the vicinity of the double eigenvalue 
(i.e., for close enough $\mu$) 
there are two close eigenvalues $\lambda$ and $(1+\epsilon)\lambda$. 
In order to find such $\lambda$ and $\mu$ we have to solve the following coupled eigenvalue equations:
\begin{eqnarray}
\left(A+\mu B\right)u & = & \lambda u\\
\left(A+\mu B\right)v & = & \lambda\left(1+\epsilon\right)v,
\end{eqnarray}
where $I$ is the $n \times n$ identity matrix. 
This kind of problem is called \textquotedblleft the two-parameter eigenvalue problem\textquotedblright. 
The most common way to solve and analyze two-parameter eigenvalue problems
is by means of three so-called matrix determinants
\begin{eqnarray}
		\Delta_{0} & = & -I\otimes B+(1+\epsilon)B\otimes I\\
		\Delta_{1} & = & -A\otimes B+B\otimes A\\
		\Delta_{2} & = & I\otimes A-(1+\epsilon)A\otimes I.
\label{eq:MFRD_matrix_det}
\end{eqnarray}
These are $n^{2}\times n^{2}$ matrices.
After constructing these matrices, we solve the following generalized eigenvalues problems:
\begin{eqnarray}
		\lambda\Delta_{0}z & = & \Delta_{1}z\\
		\mu\Delta_{0}z & = & \Delta_{2}z,
\label{eq:MFRD_genral_EV}
\end{eqnarray}
to get the approximation for $\mu$ and $\lambda$ and a tensor product $z=u\otimes v$.

The value of $\epsilon$ has to be small, in order to reflect the double eigenvalue, 
but not too small in order to maintain stability.
As a rule of thumb, a good choice is
\begin{equation} 
\epsilon\sim\epsilon_{mach}^{1/3},
\label{eq:MFRD_eps}
\end{equation}
where $\epsilon_{mach}$ is the machine precision. 

To summarize, the steps of the method follows.
Given two $n\times n$ matrices $A$ and $B$:
\begin{enumerate}
	\item Choose appropriate $\epsilon$ (see Eq. (\ref{eq:MFRD_eps})).
	For $\epsilon_{mach}=2.2\times10^{-16}$ (Matlab), we have
	 $\epsilon_{mach}^{1/3}\approx6\times10^{-6}$
	\item Construct the matrix determinants of Eq. (\ref{eq:MFRD_matrix_det}).
	\item Solve the generalized eigenvalues Eq. (\ref{eq:MFRD_genral_EV}) 
	problem to get the approximation for $\mu$ and $\lambda$.
\end{enumerate}
By construction, This method yields only an approximation to the pairs ($\lambda$,$\mu$).
But this approximation can be an initial guess for an iterative method or an exact one to get an exact pair ($\lambda$,$\mu$).

\section{Noise sensitivity of the harmonic inversion}
\label{sec:noise_sensitivity}
Parameters estimation naturally raises the issue of sensitivity to noisy experimental data.
The noise sensitivity will be determined by the method of harmonic inversion. 
If the sampling periods have high accuracy then the time series can be shown to have an underlying Hamiltonian
generator. 
This is the basis for linear methods, such as the filter diagonalization (FD) 
\cite{neuhauser1995FDM,mandelshtam2001FDM}. 
The noise in these methods results in normally distributed underlying matrices, 
and the model displays monotonous behaviour with respect to the noise.
This was verified analytically and by means of simulations In Ref. \cite{Benko2008noisyFDM}.
As a result sufficient averaging will eliminate the noise.
Practical implementations require further analysis with evidence of nonlinear effects of noise. 
For example, Mandelshtam et. al. analysed the noise-sensitivity of the FD in the context of NMR experiments 
\cite{hu199876FDM_noise,Celik2010FDM_sensitivity} 
and Fourier transform mass spectrometry \cite{martini2014mass_spectrometry}.
For some other methods, a noise reduction technique was proposed in Ref. \cite{main2000threeHI}.

\subsection*{Bibliography}
\bibliographystyle{unsrt}
\bibliography{EP_in_Bloch3_b1}

\end{document}